# Coherence-Incoherence and Dimensional Crossover in Layered Strongly Correlated Metals

T. Valla*, P. D. Johnson*, Z. Yusof†, B. Wells†, Q. Li‡, S. M. Loureiro§, R. J. Cava§, M. Mikami|| Y. Mori||, M. Yoshimura|| and T. Sasaki||

**Physics Department, Brookhaven National Laboratory, Upton, NY 11973, USA.*

*†Dept. of Physics, University of Connecticut, 2152 Hillside Road U-46, Storrs, CT 06269, USA.*

*‡ Material Sciences Department, Brookhaven National Laboratory, Upton, NY 11973, USA.*

*§Department of Chemistry and Princeton Materials Institute, Princeton University, Princeton NJ, USA.*

*||Department of Electrical Engineering, Osaka University, 2-1 Yamada-oka, Suita-shi, Osaka, 565-0871 Japan*

**Correlations between electrons and the effective dimensionality are crucial factors that shape the properties of an interacting electron system. For example, the onsite Coulomb repulsion, *U*, may inhibit, or completely block the intersite electron hopping, *t*, and depending on the ratio *U*/*t*, a material may be a metal or an insulator.[1] The correlation effects increase as the number of allowed dimensions decreases. In 3D systems, the low energy electronic states behave as quasiparticles (QP), while in 1D systems, even weak interactions break the quasiparticles into collective excitations.[2] Dimensionality is particularly important for a class of new exotic low-dimensional materials where 1D or 2D building blocks are loosely connected into a 3D whole. Small interactions between the blocks may induce a whole variety of unusual transitions. Here, we examine layered systems that in the direction perpendicular to the layers display a crossover from insulating-like ($\partial r/\partial T < 0$), at high temperatures, to metallic-like character ($\partial r/\partial T > 0$) at low temperatures, while being metallic over the whole temperature range within the layers. We show that this change in effective dimensionality correlates with the existence or non-existence of coherent quasiparticles within the layers.**

The crossover in the interlayer transport has been detected in layered metals such as $Sr_2RuO_4$ [3] and $NaCo_2O_4$ [4] and more recently in $(Bi_{1-x}Pb_x)_2M_3Co_2O_y$ (M=Ba or Sr) [5,6]. The layers appear as "isolated" at high temperatures, but connected at low temperature to give a 3D system. A similar crossover is observed in the quasi-1D Bechgaard salt $(TMTSF)_2PF_6$ in the least conductive direction.[7] The crossover temperature, $T_M$, is typically between 90 and 200 K. The low temperature phase ($T \ll T_M$) is 3D-like in the sense that the resistivity has nearly the same temperature dependence in all three directions. For two systems studied here, the crossover is shown later in Fig. 2(b) and Fig. 3(c).

The transport in the c-axis (perpendicular to the layers) of an anisotropic layered system may be coherent or incoherent. In a system showing coherent transport, the electronic states are well described in terms of the dispersing 3D momentum states in the usual band-formulation. The conductivity is determined by the scattering rate $\Gamma=1/\tau$ and the Fermi velocity in that direction. The system may be anisotropic, but with a nearly temperature-independent anisotropy ratio, i.e. $r_c(T)$ µ $r_{ab}(T)$ µ $\Gamma(T)$, where $r_{ab}$ and $r_c$ are in-plane and interlayer resistivities, respectively. In the case of incoherent transport, when the quasiparticle scattering rate is much greater than the effective interlayer hopping, the interlayer tunneling events are uncorrelated. Formally, this means that the electrons are scattered many times between successive tunneling events and $k_z$ is not a good quantum number. The conductivity is then proportional to the tunneling rate between two adjacent layers, and assuming that the intra-layer momentum is conserved in a tunneling event[8,9]:

$$s_c(T,w=0) \propto \int \frac{d^2k}{(2p)^2} t_\perp(k)^2 G_R(k,w) G_A(k,w) \frac{\partial f(w)}{\partial w} \qquad (1)$$

where $G_{R,A}$ is the retarded or advanced in-plane Green's function near the Fermi level ($w$=0) and $f$ is the Fermi distribution. In the quasiparticle picture, the Green's functions have a coherent component $\sim Z/(w-e_k-i\Gamma)$, resulting in the interlayer transport (1) having the same temperature dependence as the in-plane transport, dictated by $\Gamma(T)$. The behavior observed in Figures 2(b) and 3(c) therefore suggests that the quasiparticle picture is inappropriate above $T_M$, where conductivities are uncoupled[8].

ARPES is an ideal experimental probe for testing the character of excitations. It measures directly the same single-particle spectral function $A(k,w) \sim \mathrm{Im}\, G(k,w)$ that enters the transport equation (1). However, in contrast to transport probes, ARPES is $k$-resolving and able to probe deeper states that may contribute to the transport only at higher temperatures, $k_B T \sim w$. Applying ARPES studies to these materials we find that the crossover observed in transport strongly correlates with changes observed in the spectral function.

In Fig. 1 we show the photoemission intensity recorded from $(Bi_{0.5}Pb_{0.5})_2Ba_3Co_2O_y$ in the (0,0) to (π,π) direction of the Brillouin zone. This material is a non-superconducting relative of the double-layer cuprate superconductor $Bi_2Sr_2CaCu_2O_{8+\delta}$, where the Cu-O planes are substituted by Co-O. Resistivity measurements on samples from the same batch (Fig. 2(b)) give metallic $r_{ab}$, while $r_c$ shows a crossover at $T_M$ ~200 K. The anisotropy $r_c/r_{ab}$ increases with decreasing temperature, saturating below ~150 K at a value of ~$7\times10^3$.

The wide range, low temperature spectrum (panel (a)) shows a broad (~0.8 eV FWHM) hump, centered at ~0.6 eV, and a sharp state close to the Fermi level with a dip in between. The sharp peak disperses with $k_{||}$ and crosses the Fermi level with a small velocity $u_F$~160 meVÅ (note that this is one of the smallest Fermi velocities ever measured in ARPES), as visible in the expanded low energy region of the contour spectrum (panel (b)). Our detailed studies indicate that this observation does not depend significantly on the in-plane azimuth. The sharp state forms a large, nearly cylindrical ($k_F \approx 0.5$Å$^{-1}$), hole-like Fermi surface centered at the Γ point, which remains ungapped down to the lowest measured temperature, in accord with the metallic character of Co-O planes. The state is well defined only at low temperatures and only within the range of ~40-50 meV from the Fermi level. At higher temperatures, when $k_BT$ becomes comparable to that energy scale, the state loses coherence and the sharp peak diminishes. This is illustrated in panels (b) and (c) where the spectral function is plotted on the same scale with the Fermi distribution for two given temperatures. With increasing temperature the spectrum loses its fine structure near the Fermi level. In Fig. 2 we show the temperature development of the spectral function in more detail, and correlate it with the crossover in transport. The energy distribution curves (EDCs) for $k \approx k_F$ are shown for several temperatures. The sharp peak both loses intensity and broadens with temperature, with the final result that between 180K and 230K the spectral function, that was locally peaked at $w$=0 at low temperature, becomes a monotonically increasing function of energy at higher temperatures. The low-energy structure is lost and only the broad hump characterizes the spectrum. These changes in the spectral function correlate with the crossover temperature in c-axis resistivity, $T_M$~200 K.

The second system that we present is $NaCo_2O_4$. This material crystallizes in a layered hexagonal structure with a triangular Co lattice, octahedrally coordinated with O above and below the Co sheets. The Na ions are in the planes between the $CoO_2$ layers. The transport properties show many peculiarities,[4] including a c-axis crossover at $T_M$~180 K (Fig. 3(c)). This system is a much better conductor than the previous one, with $r_{ab} \sim 300 m\Omega\text{cm}$ and an anisotropy of ~40 at room temperature. The anisotropy increases at lower temperatures and saturates below ~120 K at a value of ~150.

The spectral function of $NaCo_2O_4$, shown in Fig. 3, shares common features with the previous system: a broad hump at high binding energy (~1.2 eV FWHM) and a sharp feature that crosses the Fermi level. At high temperatures, the sharp state again disappears and the low energy part of the spectrum is a smooth, increasing function of energy. This transformation in the spectrum again correlates with the crossover in resistivity.

Common to both materials, when the coherent quasiparticles form in the plane, the c-axis transport becomes metallic and the system behaves as an anisotropic 3-D metal. It is not clear, however, whether the in-plane coherence is a consequence or the cause of the dimensional crossover. One possibility is that the coherence occurs and the third dimension develops when the temperature becomes smaller than the effective energy scale for c-axis hopping. With all three dimensions "visible" ( $k_B T << t_i,\ i = x, y, z$ ), the low-energy excitations are quasiparticles. This is in line with the reasoning characterizing quasi-one dimensional metals, where the 1-D to 3-D crossover is dictated by the finite inter-chain ($t_\perp$) hopping integral.[10] In an alternative picture, sufficiently coherent ($\Gamma(T) << Zt_z$) QPs must first be formed in the planes to induce or allow coherent c-axis transport between the planes.

The crossover from coherent to incoherent excitations and a maximum in resistivity at a finite temperature is reminiscent of the Kondo behavior in heavy fermion systems.[11] The latter crossover is related to the appearance of coherent, strongly renormalized ($m^*/m_e$~100-1000) quasiparticle states below the temperature at which the *f*-moments order, as recently observed in a photoemission study[12]. The systems studied here are, on the other hand, close to a Mott-Hubbard type of metal to insulator transition (MIT), and in one of them, $(Bi_{1-x}Pb_x)_2Ba_3Co_2O_y$, the transition is experimentally realized by changing the carrier concentration by lead doping.[5] Recent theories of the MIT[13,14,15], suggest that the same physics responsible for Kondo behavior, also shapes the physical properties near the MIT. In the metallic regime, these theories predict a sharp peak in the density of states near the Fermi level, in addition to separated lower and upper Hubbard bands. The sharp peak corresponds to a strongly renormalized quasiparticle band whose effective mass diverges as $U/t$ increases to some critical value at the MIT. This state is responsible for the Fermi-liquid behavior at low temperatures. However, it disappears above some characteristic temperature, and the spectral function, controlled by the high-energy physics ($U$), becomes incoherent. The transport is no longer governed by coherent quasiparticles, but by collective excitations with resistivities going well beyond the Mott-Ioffe-Regel limit into a regime where the QP mean free path would be smaller than the inter-atomic distances[16].

Depending on the ratio, *U*/*t*, the incoherent response may even acquire a form characterizing the insulating side of the MIT. It is important to note that although these models capture the incoherence-coherence transition, they are isotropic and therefore inappropriate for low-dimensional systems. In anisotropic materials the different directions may be affected differently by electronic correlations. In the case of one-dimensional Hubbard chains, for example, a small inter-chain hopping may induce deconfinement and Luttinger (1D, incoherent physics) to Fermi liquid (3D, coherent excitaitons) crossover[17]. For layered metals, equation (1) formally means that *once the coherent part of the in-plane Green's function has disappeared* ($Z \rightarrow 0$), $r_c(T)$ *is no longer bound by the in-plane transport*. In other words, if the response is governed by collective excitations instead of quasiparticles, it may take different forms in different directions.

In other systems a similar dimensional crossover to a coherent, 3D low-temperature state may be induced by different mechanisms. For example, the cuprate high temperature superconductors (HTSCs) near optimal doping have metallic planes with an insulating inter-layer resistivity, again suggesting that the normal state transport is a collective phenomenon[8]. Indeed, the spectral function shows the absence of sharp quasiparticle peaks in the normal state. However, these systems become (anisotropic) 3D superconductors and the quasiparticles reappear below $T_C$ in analogy with the dimensional crossovers discussed above. The transition goes directly from 2D-like normal state into the quasiparticle-like 3D superconducting state and cannot be dictated by the single-particle tunneling. In the stripe picture, for example, the normal state is considered in terms of the 1D physics within the charge stripes and the superconducting transition is viewed as a 1D to 3D transition, induced by the inter-stripe Josephson tunneling of cooper pairs that already exist within the stripes in the normal state.[18]

In an alternative scenario, a different behavior for in- and out-of -plane transport could occur if both the interlayer hopping $t_{\perp}$, and the FS in the plane are anisotropic.[9] Indeed, it has been shown[19] that in HTSCs, the scattering rate in the corners of the Brillouin Zone that dominate c-axis hopping behaves differently from the scattering rate in the nodal regions[20] where $t_{\perp}=0$. In addition, the corners of the BZ are affected by the normal state pseudogap[21] for $T_C<T<T^*$, ($T^*$ is the pseudogap temperature) that blocks the normal state c-axis transport. The inter-layer Josephson coupling then enables the transition into the 3D superconducting state without ever going through the 3D Fermi-liquid-like normal state. In the highly overdoped regime, the pseudogap disappears, de-confining the charges from the planes[22] ($\partial r_c / \partial T > 0$) even in the normal state. This transition into a 3D-like state, whether induced by single particle tunneling (normal state) or Josephson coupling (superconducting state), is accompanied by the appearance of a sharp peak in the spectral function,[23] an indication that the 3D state approaches the FL regime. The crossover may also proceed via the exchange coupling $J_{\perp}$, leading to an (anti)ferromagnetic 3D ground state. In cuprates, the low-doping regime is particularly interesting, because there are indications[24] that different mechanisms may compete in bringing the system into the 3D ground state.

1. Mott, N., F., in Metal-Insulator Transitions. (Taylor & Francis, London, New York, c1990).

2. Jâerome, D. & Caron, L., G., in Low-dimensional conductors and superconductors. (Plenum Press, New York, c1987).

3. Maeno, Y. *et al*, Superconductivity in a layered perovskite without copper. *Nature* **372**, 532-534 (1994).

4. Terasaki, I. Sasago, Y. & Uchinokura, K., Large thermoelectric power in $NaCo_2O_4$ single crystals. *Phys. Rev. B* **56**, R12685-R12687 (1997).

5. Loureiro, S. M. *et al*, Enhancement of metallic behavior in bismuth cobaltates through lead doping. *Phys. Rev. B* **63**, 094109(1-9) (2001).

6. Tsukada, I. *et al*, Ferromagnetism and large negative magnetoresistance in Pb doped Bi-Sr-Co-O misfit-layer compound. *cond-mat/0012395*, (2000).

7. Mihaly, G. Kezsmarki, I. Zambroszky, F. & Forro, L. Hall Effect and Conduction Anisotropy in the Organic Conductor (TMTSF)2PF6. *Phys. Rev. Lett.* **84**, 2670-2673 (2000).

8. Anderson, P., W., in The Theory of Superconductivity in the High-$T_c$ Cuprates. (Princeton Univ. Press, Princeton, NJ, 1997).

9. Ioffe, L., B. & Millis, A., J., Zone-Diagonal-Dominated Transport in High-TC Cuprates. *Phys. Rev. B* **58**, 11631-11637 (1998).

10. Voit, J. One-Dimensional Fermi Liquids. *Rep. Prog. Phys.* **57**, 977-1116 (1994).

11. Fisk, Z. *et al*. Heavy-Electron Metals: New Highly Correlated States of Matter. *Science* **239**, 33-42 (1988).

12. Reinert, F. *et al*. Temperature Dependence of the Kondo Resonance and Its Satellites in $CeCu_2Si_2$. *Phys. Rev. Lett.* **87**, 106401(1-4) (2001).

13. Pruschke, T., Cox, D., L. & Jarrell, M. Hubbard Model at Infinite Dimensions: Thermodynamic and Transport Properties. *Phys. Rev. B* **47**, 3553-3565 (1993).

14. Georges, A., Kotliar, G., Krauth, W. & Rozenberg, M. J. Dynamical Mean-Field Theory of Strongly Correlated Fermion Systems and the Limit of Infinite Dimensions. *Rev. Mod. Phys.* **68**, 13-125 (1996).

15. Merino, J. & McKenzie, R., H. Transport Properties of Strongly Correlated Metals: A Dynamical Mean-Field Approach. *Phys. Rev. B* **61**, 7996-8008 (2000).

16. Ioffe, A., F. & Regel, A., R. Non-crystalline, amorphous and liquid electronic semiconductors. *Prog. Semicond.* **4**, 237-291 (1960).

17. Biermann, S., Georges, A., Lichenstein, A. & Gianmarchi, T. Deconfinement Transition and Luttinger to Fermi Liquid Crossover in Quasi One-Dimensional Systems. *cond-mat/0107633* (2001).

18. Carlson, E., W., Orgad, D., Kivelson, S., A. & Emery, V., J. Dimensional Crossover in Quasi-One-Dimensional and High-$T_C$ Superconductors. *Phys. Rev. B* **62**, 3422-3437 (2000).

19. Valla, T. *et al*. Temperature Dependent Scattering Rates at the Fermi Surface of Optimally Doped $Bi_2Sr_2CaCu_2O_{8+\delta}$. *Phys. Rev. Lett.* **85**, 828-831 (2000).

20. Valla, T. *et al*. Evidence for Quantum Critical Behavior in the Optimally Doped Cuprate $Bi_2Sr_2CaCu_2O_{8+\delta}$. *Science* **285**, 2110-2113 (1999).

21. Ding, H. *et al*, Spectroscopic Evidence for a Pseudogap in the Normal State of Underdoped High-$T_C$ Superconductors. *Nature* **382**, 51-54 (1996).

22. Watanabe, T., Fujii, T., & Matsuda, A., Pseudogap in $Bi_2Sr_2CaCu_2O_{8+\delta}$ Studied by Measuring Anisotropic Susceptibilities and Out-of-Plane Transport. *Phys. Rev. Lett*. **84**, 5849-5852 (2000).

23. Yusof, Z. *et al*, Quasiparticle Liquid in the Highly Overdoped $Bi_2Sr_2CaCu_2O_{8+\delta}$. *cond-mat/0104367* (2001).

24. Lavrov, A. N., Kameneva, M. Y. & Kozeeva, L. P., Normal-State Resistivity Anisotropy in Underdoped $RBa_2Cu_3O_{6+x}$ Crystals. *Phys. Rev. Lett.* **81**, 5636-5639 (1998).

We acknowledge A. Tsvelik, R. Werner, S. A. Kivelson and A. V. Fedorov for useful discussions. The work at BNL and the National Synchrotron Light Source (NSLS) where the experiments were carried out are supported by the U.S. Department of Energy.

**Correspondence and requests for materials should be addressed to T.V. (e-mail: valla@bnl.gov)**

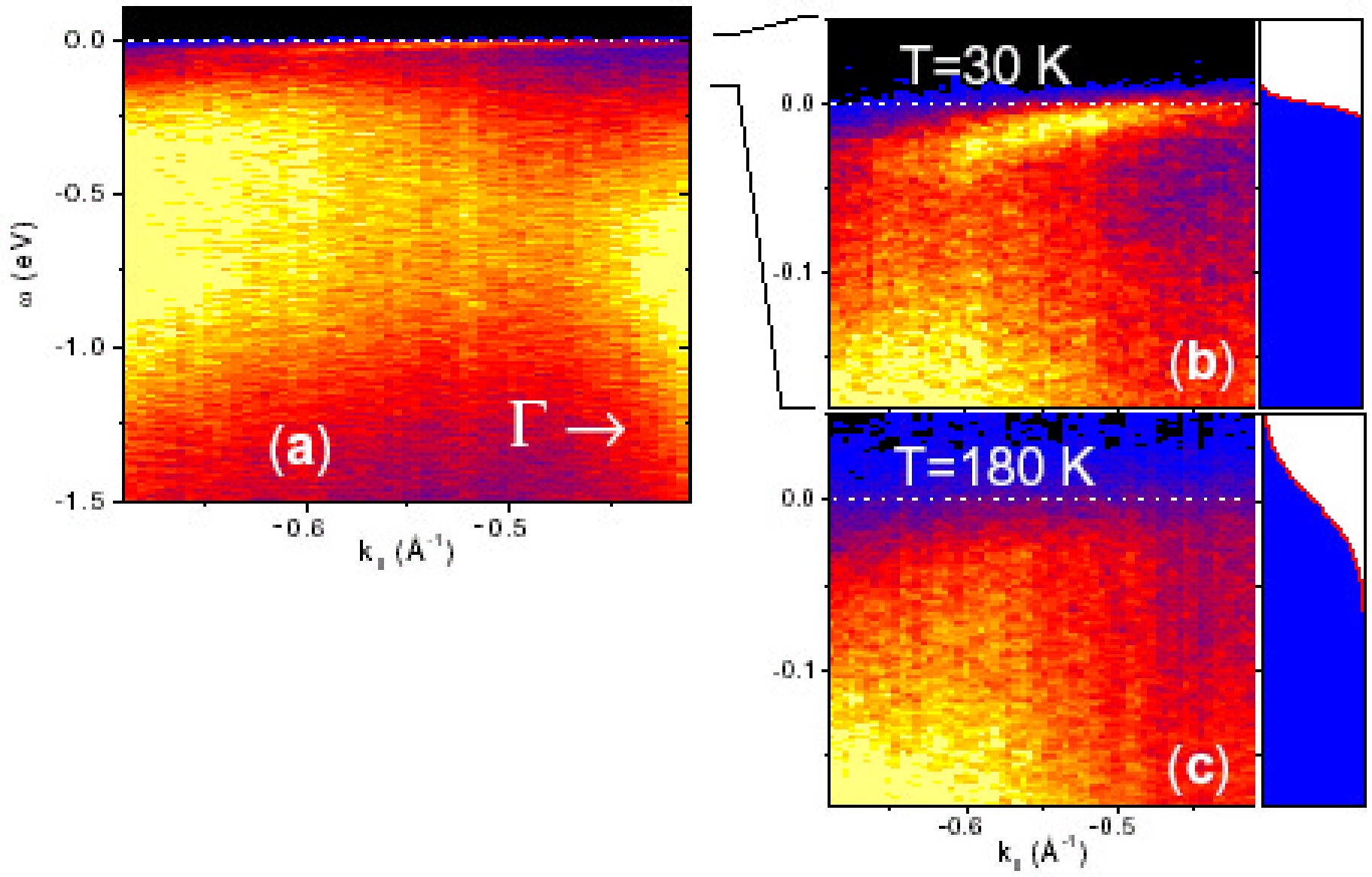

Figure 1. ARPES from $(Bi_{0.5}Pb_{0.5})_2Ba_3Co_2O_y$. The contour plot of the ARPES intensity along the (0,0) to (π,π) line at T=30 K (panel (**a**)). The arrow indicates the direction to the zone center $\bar{\Gamma}$. Changes in the low-energy region with temperature: the 30 K (panel (**b**)) and 180 K (panel (**c**)) contour spectra (left) are plotted on the same scale with the corresponding Fermi distributions (right). The spectra were taken with a Scienta SES200 hemispherical analyzer with an angular resolution of $\pm 0.1^o$ or better and an energy resolution of the order of 10 meV. Photons of 15.4 eV, provided by a Normal Incidence Monochromator based at the NSLS, were used for the excitation. Samples, grown as described in ref. [5], were mounted on a liquid He cryostat and cleaved *in situ* in the ultra-high vacuum chamber with base pressure $2\text{-}3\times10^{-9}$ Pa.

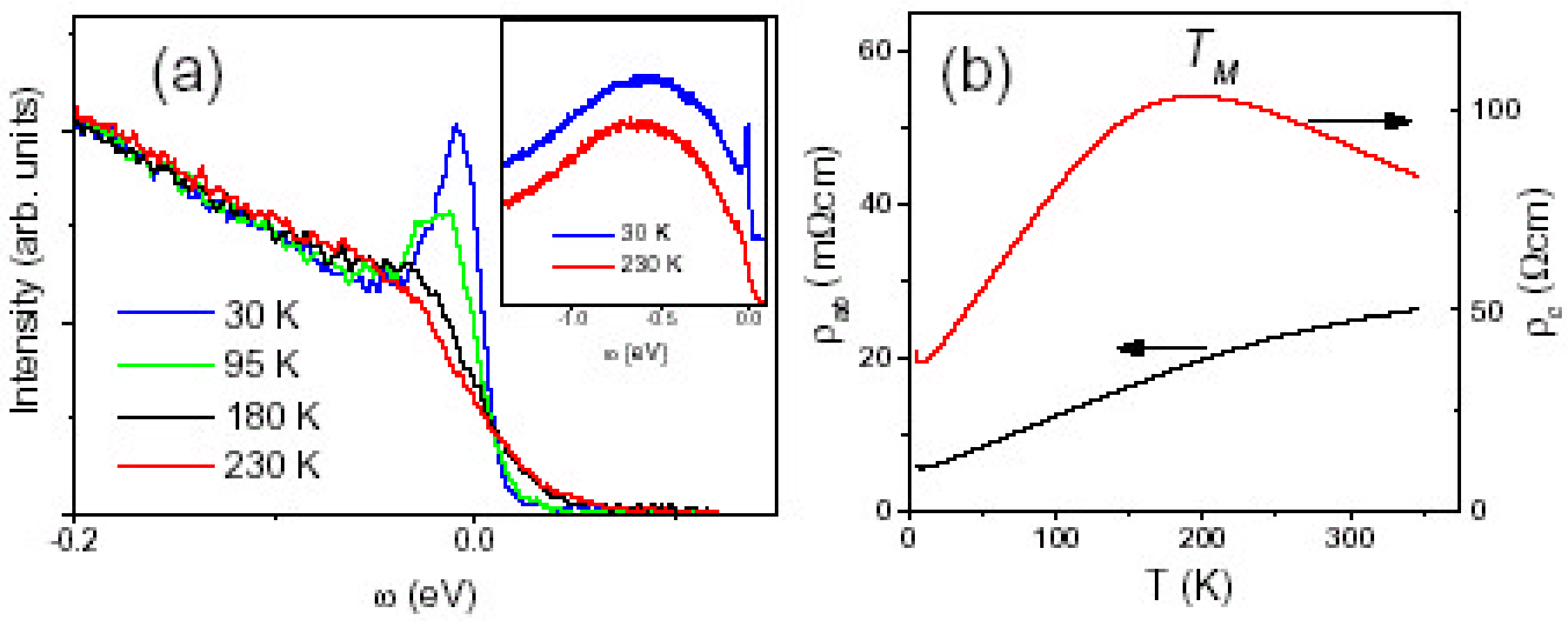

Figure 2. Correlation between the ARPES and transport in $(Bi_{0.5}Pb_{0.5})_2Ba_3Co_2O_y$. The changes in EDCs (for $k$»$k_F$) with temperature are shown in panel (a). The inset shows the wide-range EDCs. The in- and the out-of-plane resistivities (panel (b)) are measured on a sample from the same batch with a conventional four-probe technique.

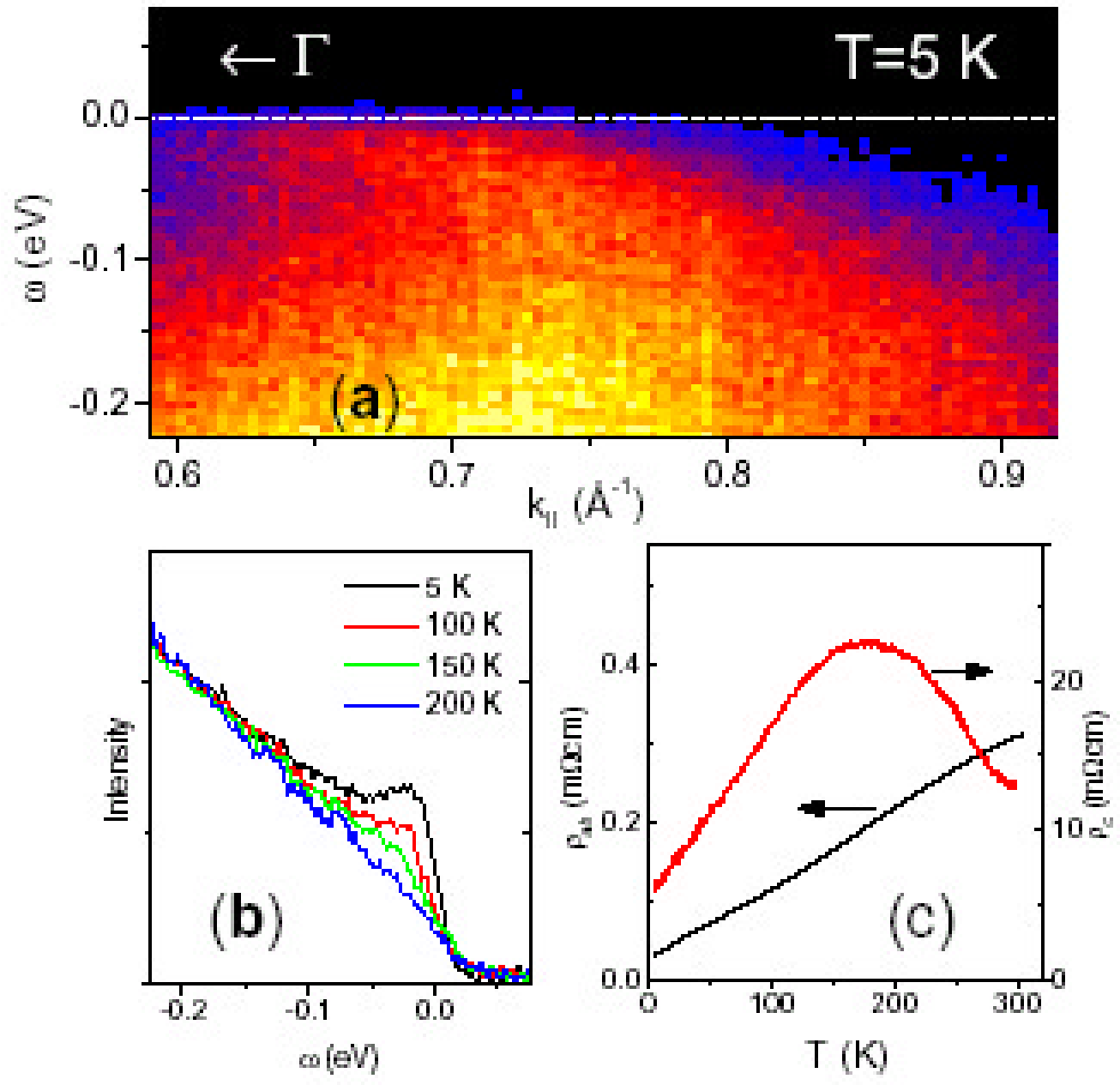

Figure 3. ARPES and dc-transport in $NaCo_2O_4$. The contour plot (panel (**a**)) shows the ARPES intensity along the $\overline{\Gamma}$ - $\overline{K}$ line of the Brillouin zone at T=5K. The Fermi crossing is a point on a large hole-like Fermi surface ($k_F$~0.7Å$^{-1}$) centered at the Γ point. The photon energy was 21.22 eV (HeI radiation). The samples are NaCl based flux grown. The temperature dependence of the low-energy region of EDCs (for $k$»$k_F$) is shown in panel (**b**). The in- and the out-of-plane resistivities (panel (c)) are measured on a sample from the same batch.